\documentclass[article]{revtex4}

\usepackage{graphicx}                                  
\usepackage{epstopdf}                                  
\usepackage{dcolumn}                                   
\usepackage{bm}                                  
\usepackage{amssymb}                                   
\usepackage{amsmath}                                   
\usepackage[english]{babel}                            
\usepackage[usenames, dvipsnames]{color}               
\usepackage{mathtools}                                 
\usepackage{mathrsfs}

\begin{document}
	
	
	\title{Temperature dependence of resistivity at the transition to a charge density wave state in rare-earth tritellurides}
	
	\author{K.K. Kesharpu}
	\email{m1708377@edu.misis.ru}
	\affiliation{Department of Theoretical Physics \& Quantum Technology,National University of Science and Technology MISiS, 119049, Moscow, Russia}
	\author{P.D.\ Grigoriev}
	\email{grigorev@itp.ac.ru}
	\affiliation{L.D. Landau Institute for Theoretical Physics, 142432, Chernogolovka, Russia}
	\affiliation{Department of Theoretical Physics \& Quantum Technology,National University of Science and Technology MISiS, 119049, Moscow, Russia}
	\affiliation{P.N. Lebedev Physical Institute, RAS, 119991, Moscow, Russia}
	\date{\today }

	
	\begin{abstract}
About a half of the Fermi surface in rare-earth tritellurides $RTe_3$ becomes gapped below the transition to a charge-density-wave (CDW) state, as revealed by ARPES data. However, the observed jump in resistivity during the CDW transition is less than $20$\%. Previously this phenomenon was explained by hypothesizing a very slow evolution of \textit{CDW} energy gap below transition temperature in $RTe_3$ compounds, which contradicts the X-ray measurements. Here we show that this weak change in resistivity can be explained in the framework of standard mean-field temperature dependence of the CDW energy gap in agreement with X-ray data. The change of resistivity caused by CDW is weak because the decrease in conducting electron density at the Fermi level is almost compensated by the decrease in their scattering rate. We calculate resistivity in $RTe_3$ compounds using the Boltzmann transport equation and the mean-field description of the CDW state, and obtain a good agreement with experimental data.
	\end{abstract}
\maketitle
\date{\today}
	
	
	\section{Introduction}	
	
The charge-density-wave (CDW) ground state is rather common in strongly anisotropic metals \cite{Gruner1994,Monceau12}. The typical precursors of the CDW transition are strong electron-phonon (EPC) or/and electron-electron (e-e) coupling and the nesting of Fermi surface (FS),\cite{Gruner1994} though the latter is not always a determining factor\cite{Mazin2008,Zhu2017a}. In metals with (almost) ideal FS nesting the CDW creates an energy gap on the Fermi level and converts a metal to an insulator or semiconductor. In most CDW compounds the FS nesting is not perfect; then the FS is only partially gapped in the CDW state, and the compound retains its metallic properties even in the CDW state. An example of such partially gapped CDW compounds is the family of quasi-2D layered material, known as \textit{rare earth tritelluride ($RTe_3$ :R=Y, La, Ce, Nd, Sm, Gd, Tb, Ho, Dy, Er, Tm)} \cite{Norling1966}. These compounds are very convenient for the demonstration of various electronic properties in a partial CDW-metallic state, and their electronic structure was studied extensively and with rather high precision using {\it ARPES (Angle Resolved Photo Emission Spectroscopy)} \cite{Brouet2008,Brouet2004,Schmitt2011} and through various other probing methods \cite{Lavagnini2008,Ru2008,Ru2008a,Sacchetti2006,Sinchenko2014,Sinchenko2014a}. 

$RTe_3$ compounds have weakly orthorhombic layered crystal structure \textit{(Space Group: Cmcm)} formed by sandwiching two alternate puckered R-Te layers in between two double Te planes. We will mainly concentrate on terbium tritelluride ($TbTe_3$), which has two incommensurate CDW: one at high temperature with $T_{c1}=336K$ and the second below much lower temperature $T_{c2}=41K$. Above $T_{c2}=41K$ it was shown to have unidirectional stripes as opposed to checkerboard pattern observed below $T_{c2}=41K$, which somewhat resembles the behavior of under-doped cuprates\cite{LeBoeuf2013}. The FS in $TbTe_3$ at two different temperatures was measured by ARPES in Ref. \cite{Schmitt2011}, and the temperature dependence of resistivity along different directions has been investigated in Refs. \cite{Sinchenko2014,Sinchenko2014a}. Experimentally it has been found that the resistivity in $TbTe_3$ shows a too small and very anisotropic jump at the transition temperature $T_{c1}$ from metallic to CDW state (see figure 1 of Ref. \cite{Sinchenko2014}), which is in contrast to the expected much larger jump. This phenomenon has been explained in Ref. \cite{Sinchenko2014} only by assuming a very weak temperature dependence of the CDW energy $\Delta (T)$ just below $T_{c1}$, attributed to strong fluctuations and given by Eq. (\ref{eqn_energy_gap}) below with $\alpha\approx 2$ instead of the mean-field-theory value $\alpha =1/2$. However, the X-ray studies\cite{Ru2008a} suggest nearly the mean-field temperature evolution of the CDW order parameter below $T_{c1}$ with $\alpha \approx 1/2$ (see Fig. 6 of Ref. \cite{Ru2008a}). In the present work we resolve this inconsistency and show that the observed T-dependence of resistivity can be explained in the framework of mean-field dependence $\Delta (T)$. 
	
	\section{Theory}	
	
The anisotropic resistivity ($\rho_{i} = 1/\sigma_{i}$) is calculated by using the {\it Boltzmann transport equation} (see Eq. (3.16) of Ref. \cite{Abriskov}), which gives
\begin{equation}
\begin{aligned}
\sigma_i(T) =&2e^{2}\sum_{\boldsymbol{k}%
}\,\tau v_{i}^{2}\left( \boldsymbol{k}\right) \left[ -\partial f_0/\partial \varepsilon \right] \\&= 
- e^2 \int v_{i}^2 \tau \frac{\partial f_0}{\partial \varepsilon} \: g(\varepsilon) d\varepsilon \frac{d\Omega}{4 \pi}\: ,
\label{Conductivity_eqn}
\end{aligned}
\end{equation}
where $\sigma_{i}(T)=1/\rho_{i}$ is the conductivity along main axes $i=x,y,z$, $v_i(k)$ is electron velocity, $\boldsymbol{k}=\{k_x,k_y,k_z\}$ is the vector of electron momentum, $\partial  f_0/ \partial \varepsilon =-1/\{4T\cosh^{2}\left[(\varepsilon-E_F)/(2T)\right]\}$ is the derivative of the Fermi distribution function, which restricts the summation over momentum to the vicinity of FS, $E_F$ is the Fermi energy, $g(\varepsilon)$ is the {\it density of states (DOS)}, which depends on energy $\varepsilon $, and $d\Omega$ is the solid angle extended by an area in momentum space. $\tau$ is the mean scattering time. At $T\sim T_{c1}=336K\gg T_D$, where $T_D\approx 180$K is the Debye temperature, the electron scattering comes mainly from the short-wavelength phonons. This scattering is similar to the scattering by short-range impurites, because it also gives the momentum transfer of the order of Brillouin zone length, while the energy transfer is $\sim k_B  T_D\ll T$. Hence, similarly to the scattering by short-range impurities, in the Born approximation or according to the golden Fermi rule, the scattering rate $1/\tau$ is proportional to the density of electron states $g(\varepsilon )$ \cite{Abriskov}, which has a jump at $T_{c1}$ due to the opening of CDW energy gap. Also $1/\tau$ is proportional to the number of phonons, which grows linearly with $T$. Hence, the temperature dependence of scattering time is approximately given by $\tau (T) \propto [T\rho (\varepsilon, T)]^{-1}$. Substituting this to Eq. (\ref{Conductivity_eqn}) we can hypothesize that, in the CDW state as the temperature decreases, the reduction of the density of electron states contributing to conductivity is compensated by the increase in their scattering time. Hence, one does not observe a large jump in resistivity during the transition between CDW and metallic state near $T_{c1}$. Below we substantiate this idea by direct calculations.

For materials with electron dispersion (ED) $\varepsilon(k_x,k_y,k_z,T)$ the DOS is calculated as
\begin{equation}\label{eqn_dos_general_state_3D}
\begin{split}
&g(\varepsilon) = \frac{2}{(2\pi)^3}\iiint dk_x dk_y dk_z \: \delta[E-\varepsilon(k_x,k_y,k_z,T)] \:  ,
\end{split}
\end{equation}
where the factor 2 is due to spin degeneracy of electrons. The integration in Eq. (\ref{eqn_dos_general_state_3D}) is over the first Brillouin zone, but due to the $\delta $-function it is restricted to the FS, which is found by solving the equation $\varepsilon(k_x,k_y,k_z,T)=E_F$. 

Because we are interested in layered materials, all the analysis is done for 2D case (although the same procedure can be applied to 3D materials). For 2D material the electron dispersion depends only on 2 momentum components (let it be $k_x$ and $k_y$). Then Eq. (\ref{eqn_dos_general_state_3D}) reduces to
\begin{equation}\label{eqn_dos_general_state_2D}
\begin{split}
&g(\varepsilon) = \frac{2}{(2\pi)^2}\iint dk_x dk_y\: \delta[E-\varepsilon(k_x,k_y,T)] \:  .
\end{split}
\end{equation}
Below all functions in the text explicitly depends on $k_x$, $k_y$, $T$ unless otherwise mentioned.

In the mean field approximation the electron dispersion (ED) in a CDW state is given by\cite{Gruner1994}
\begin{equation}\label{eqn_full_electronic_dispersion}
\begin{aligned}
\varepsilon(k) = \frac{\epsilon(k)+\epsilon(k-Q)} {2} \pm \sqrt{\frac{\left[\epsilon(k) - \epsilon(k-Q)\right]^2}{4} +\Delta_0^2(T)} \: ,
\end{aligned}
\end{equation}
where Q is the CDW wave vector and the energy gap $\Delta$ is momentum-independent. If the anti-nesting term $\epsilon(k)+\epsilon(k-Q)<2\Delta_0(T)$, the corresponding $k$-state acquires a gap at the Fermi level. The mean-field dispersion in Eq. (\ref{eqn_full_electronic_dispersion}) is often simplified to
\begin{equation}
\begin{aligned}
\varepsilon(k) \approx \sqrt{\frac{\left[\epsilon(k) - \epsilon(k-Q)\right]^2}{4} +\Delta^2(k,T)}\: ,
\end{aligned}\label{eqn_reduced_electronic_dispersion}
\end{equation}
where the energy gap $\Delta (k,T)$ now depends on electron momentum and reduces with the increase of the anti-nesting term $\epsilon(k)+\epsilon(k-Q)$, being non-zero only if $\epsilon(k)+\epsilon(k-Q)<2\Delta_0(T)$. Since for the well-nested parts of FS $ \left[\epsilon(k) - \epsilon(k-Q)\right]/2\approx \epsilon(k)$, Eq. (\ref{eqn_reduced_electronic_dispersion}) simplifies to \begin{equation}
\varepsilon(k) \approx \sqrt{\epsilon(k)^2 + \Delta(k,T)^2} .
\label{eqn_reduced_electronic_dispersionF}
\end{equation} 
We will use this reduced ED for further calculations.

In $RTe_3$ materials in the CDW phase the FS is only partially gapped, i.e. for wave vector $|k_x|<k_{x0}$ the ED contains nonzero energy gap $\Delta$, and for the remaining part $\Delta =0$. Hence, the ED for $|k_x|<k_{x0}$ is given as $\varepsilon(k) = \sqrt{\epsilon(k)^2 + \Delta(T)^2}$, and for the remaining part of FS it transforms to $\varepsilon(k) = \epsilon(k)$. Due to this discontinuity in ED, the DOS $g(\varepsilon)$ for partially gapped FS is calculated piecewise. Hence, $g(\varepsilon)$ from Eq. (\ref{eqn_dos_general_state_2D}) for the partially gapped FS is found as
\begin{equation}\label{eqn_dos_cdw_state}
\begin{aligned}
g_{tot}(\varepsilon)&=g_{gap}(\varepsilon)+g_{ungap}(\varepsilon)\\
&=\frac{1}{\pi^2}\int_{-\pi /a}^{\pi /a} dk_y  \int_{0}^{k_{x0}}dk_x\, \delta (\varepsilon -\sqrt{\epsilon^2+\Delta^2}) \\
+&\int_{k_{x0}}^{\pi/a}dk_x\, \delta (\varepsilon-\epsilon)
\end{aligned}
\end{equation}
Where $g_{tot}(\varepsilon)$ is the total DOS in the CDW state, $g_{gap}(\varepsilon)$ is the DOS contribution from the gapped parts of FS,  $g_{ungap}(\varepsilon)$ is DOS from the ungapped part of FS, $a$ is the lattice constant. 
We should note that Eq. (\ref{eqn_dos_cdw_state}) is temperature dependent. Above $T_{c1}$ the first term in Eq. (\ref{eqn_dos_cdw_state}) vanishes and total DOS is only given by ungapped part, which is the metallic DOS $g_{met}(\varepsilon)$. 
	
For $RTe_3$ compounds $\epsilon$ is found in the tight-binding approximation (see Eq. (2) of Ref. \cite{Brouet2008}):
\begin{equation}
	\begin{aligned}
	\epsilon = -2t_{\parallel}\cos[(k_x \pm k_y)a/2]
	-2t_{\perp}\cos[(k_x \mp k_y)a/2] - E_F \quad ,
	\end{aligned}\label{eqn_electronic_dispersion}
\end{equation}
where $t_{\parallel}$ and $t_{\perp}$ are the parallel- and perpendicular-to-chain electron hopping integrals. Momentum dependence of energy gap $\Delta$ is found by fitting the experimental data (see Fig. 13 of Ref. \cite{Brouet2008}):
\begin{equation}
	\begin{aligned}
	 \Delta = \Delta_0 \left( 1-\frac{T^2}{T_{c1}^2}\right)^\alpha \left(1-\frac{k_x^2}{k_{xo}^2} \right) \:  ,
	\end{aligned}\label{eqn_energy_gap}
\end{equation}
In the mean-field approximation the temperature dependence of $\Delta$ is given by $\alpha=1/2$, which we use in our calculations. If $t_{\parallel} \gg t_{\perp}$ the ED in Eq. (\ref{eqn_electronic_dispersion}) simplifies to a quasi-1D chain ED, given by Eq. (1.13) of Ref. \cite{Gruner1994}.
	
 The procedure of calculating the gapped part of DOS $g(\varepsilon)_{gap}$ taking into account CDW fluctuations has already been described in Ref. \cite{Lee1973,Scalapino1972,Suzumura1987}. Here we briefly repeat the necessary steps of calculation for completeness of this article. First we find the temperature dependent coherence length $\xi$ using the {\it Ginzburg-Landau functional} (GLF) and mean-field theory. In the next step we use $\xi$ and the energy gap $\Delta$ in the one particle Green's function. The major property of this Green's function is that it considers the effect of only near neighbor degenerate states. Using this Green's function we find the electronic properties of the material. The GLF is given as
\begin{equation}
\begin{aligned}
	F &= F(0) + \int dx \left[ a |\Delta|^2 + b |\Delta|^4 + c \left|\frac{d \Delta}{d x}\right|^2\right] ,\\
	&\text{where}\\
	&a=\frac{D_0 (T-T_c)}{T_c} \quad ;\quad
	b=D_0\left[b_0+(b_1-b_0)\frac{T}{T_c}\right],\\
	&c=D_0 + \xi^2(T) \quad , \quad 
	\xi^2 = \frac{7 \zeta(3) v_f^2}{16 \pi^2 k^2 T^2} ,\\
	&b_0 = 0.5/(1.76^2) \quad , \quad 
	b_1 = \frac{7 \zeta(3)}{16 \pi^2},\\
	&D_0= \text{Density of states at $T=T_c$ at Energy $E_f .$}
\end{aligned}\label{eqn_GLF}
\end{equation}
Here the values of a, b, c are found from the mean field calculation (see Sec. 5.1 of Ref. \cite{Gruner1994} for complete derivation of the values given in Eq. (\ref{eqn_GLF})). In next step we use $\xi$ and $\Delta$ in one particle Green's functions and keep the terms with coupling between nearly degenerate states (See Eq. 5 of Ref. \cite{Lee1973}), which is written as 
\begin{equation}
	\begin{aligned}
G^{-1} (k,E) = E - &\epsilon(k) - \int dQ \: S(Q) \langle\Delta^2 \rangle\\
&\times [E - \epsilon(k\pm Q) + i\delta]^{-1}.
	\end{aligned}\label{eqn_green_function}
\end{equation}
The structure factor $S(Q)$ is the Lorentzian with origin at $Q=2k_f$ and of width $\xi^{-1}(T)$. Next we take the integral over Q in Eq. (\ref{eqn_green_function}), which results in
\begin{equation}
\begin{aligned}
G^{-1} (k,E) = E - &\epsilon(k) -  \langle \Delta^2 \rangle\\
&\times [E - v_f(|k|-k_f) + \iota v_f \xi^{-1}(T)]^{-1}.
\end{aligned}\label{eqn_green_function_integral}
\end{equation}
In the last step we integrate the imaginary part of Eq. (\ref{eqn_green_function_integral}) to get the DOS from the gapped part of FS as follows
\begin{equation}\label{gapped_dos}
\begin{aligned}
&\frac{g_{gap}(E)}{g_{met}(E)} = \frac{\alpha \sqrt{\left[2(y+x)\right]}}{\left[2(y+x)-\alpha^2\right]y},\\
&\text{where,}\\
&\tilde{\omega}= E/\Delta, \:  \:  \: 	\alpha=v_{f}/(\xi^{-1}\Delta),\\
&x=1+0.25\alpha^2-\tilde{\omega}^2, \:  y=\sqrt{(x^2 + \tilde{\omega}^2 \alpha^2)}.
\end{aligned}
\end{equation}
In Eq. (\ref{gapped_dos}) $g_{met}(E)$ is DOS in the metallic state above CDW transition temperature. 
	
The total DOS can be found as a sum of contributions $g_{gap}(\varepsilon)$ from Eq. (\ref{gapped_dos}) and $g_{ungap}(\varepsilon)$ from ungapped parts: 
\begin{equation}\label{total_dos}
\begin{aligned}
 g_{tot}(\varepsilon) = g_{met}(\varepsilon)[(1-\eta) + \eta \: g_{gap}(\varepsilon)/g_{met}(\varepsilon)] \:  ,
\end{aligned}
\end{equation}
where $\eta\approx k_{x0}a/\pi $ is the ratio between gapped and total FS length. Finally, from Eqs. (\ref{gapped_dos}),(\ref{total_dos}) we obtain
\begin{equation}\label{eqn_total_dos_in_terms_energy}
	\begin{aligned}
g_{tot}(\varepsilon) = g_{met}(\varepsilon) \left[\left(1-\frac{a k_{x0}}{\pi}\right) +\left( \frac{a k_{x0}}{\pi} \frac{\alpha \sqrt{\left[2(y+x)\right]}}{\left[2(y+x)-\alpha^2\right]y}\right) \right] \: .
	\end{aligned}
\end{equation}
\begin{figure}[h]
\includegraphics[width=18pc,height=13pc]{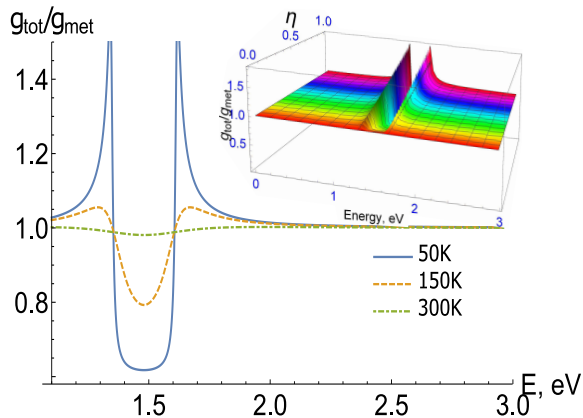}
\caption{Plot of ratio of DOS under CDW state to DOS in normal metallic state calculated from Eq. \ref{total_dos} for T=50K (blue solid line), T=150K(orange dashed line), T=300K (green dot-dashed line). $E_F =1.48$eV, $\eta =0.4$ has been used from Ref. \cite{Sinchenko2014}. Inset: The DOS dependence on $\eta$ ($0<\eta<1$) at T=50K.}
\label{plot_dos}
\end{figure}

\begin{figure}
\includegraphics[width=18pc,height=12pc]{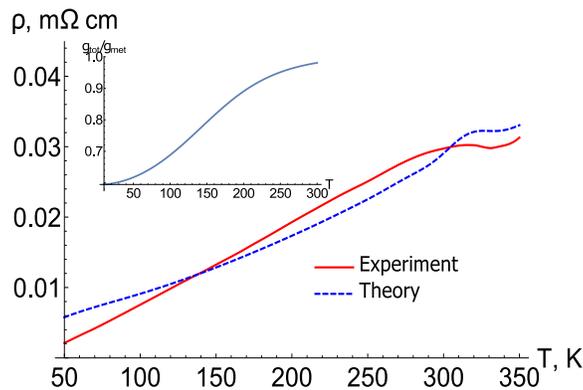}
\caption{Comparison of the experimental resistivity in $TbTe_3$ (solid red curve), extracted from Fig. 1 of Ref. \cite{Sinchenko2014}, and theoretical resistivity (blue dashed curve) calculated from Eq. \ref{Conductivity_eqn}. Experimental plot is found by taking average of $\rho_a$ and $\rho_c$.}
\label{resistivity_plot}
\end{figure}

\section{Comparison with experiment and discussion}

The ratio of DOS in CDW and metallic phases in $TbTe_3$ for three different temperatures is plotted in figure \ref{plot_dos}. In this plot we used the experimental values for $\Delta_0 =0.27 eV$, $k_{x0}=0.29 \AA^{-1}$, $\eta=0.4$, $t_{\parallel}=2$eV, $t_{\perp}=0.37$eV, $a=4.4 \AA$, $E_F =1.48$eV for $TbTe_3$ given in Ref. \cite{Brouet2008}. As expected, the DOS does not get to zero at the Fermi level for very low temperature due to contribution from the ungapped part of FS. In the inset figure 1 we show the evolution of DOS with the increase of fraction $\eta$ of the gapped part of Fermi surface. We notice that as $\eta$ goes towards unity, i.e the whole FS becomes gapped, DOS goes to zero at the Fermi level. 

The conductivity is calculated using Eq. (\ref{Conductivity_eqn}). The electron velocity $v_i$ is found by differentiating the energy dispersion relation in Eqs. (\ref{eqn_reduced_electronic_dispersionF}),(\ref{eqn_electronic_dispersion}): $v_i=\partial \epsilon / \partial k_i$. The temperature-dependent relaxation time decreases with the increase of temperature according to $\tau (T) \propto [T\rho (\varepsilon, T)]^{-1}$.
Eq. (\ref{Conductivity_eqn}) is evaluated numerically by integrating over the whole Fermi surface, and the resulting resistivity plot is shown in Fig. \ref{resistivity_plot}. As one can see from this figure, the jump of resistivity is very small at the transition temperature of 330K. This supports our hypothesis that although the number of charged quasiparticles contributing to conductivity below the CDW transition temperature decreases, it is compensated by the increase in their relaxation time $\tau $.

  The obtained theoretical curve for resistivity shows good agreement with experimental data. However, its calculation is done for the reduced electron dispersion in CDW state given by Eq. (\ref{eqn_reduced_electronic_dispersionF}). It would be useful to perform similar calculation for the full mean-field dispersion in the presence of CDW with imperfect nesting, given by Eq. (\ref{eqn_full_electronic_dispersion}). 
  Also it will be interesting to compare the results with and without the effect of fluctuations. We leave these and other relevant questions for future work. 
  
  	\section{Acknowledgment}
  	
   The work was carried out with financial support from the Ministry of Education and Science of the Russian Federation in the framework of increase Competitiveness Program of NUST ‘‘MISIS”, implemented by a governmental decree dated 16th of March 2013, No 211, and from Russian Science Foundation (project \# 16-42-01100). P.G. also thanks RFBR grant \# 17-52-150007.

\end{document}